\documentclass[11pt,a4paper, 5p, floatfix,groupedaddress, preprint]{elsarticle}
\usepackage{amsmath}
\usepackage{amssymb}
\usepackage{amsfonts}
\usepackage{mathtools}
\usepackage{graphicx}
\usepackage{url}
\usepackage[version=3]{mhchem}
\usepackage[format=plain,justification=raggedright, labelfont=bf, font=small]{caption}
\usepackage[format=plain,justification=raggedright, labelfont=bf, font=small]{subcaption}
\usepackage[english]{babel}
\usepackage{charter}
\usepackage[T1]{fontenc}
\usepackage[usenames,dvipsnames]{xcolor}
\usepackage{booktabs}
\usepackage{nicefrac}
\usepackage{tikz}
\usetikzlibrary{shapes,snakes}
\usepackage{wrapfig}
\usepackage{float}
\usepackage[section]{placeins}
\usepackage{siunitx}
\usepackage[colorlinks,allcolors=red]{hyperref}
\usepackage{multirow}

\newcommand {\dx} {\; \mathrm{d} }              

\newlength{\thirdlinewidth}
\graphicspath{{./figures/}}

\begin{document}
\begin{frontmatter}
\title{\Large 2D foams above the jamming transition: Deformation matters}

\author[TCD]{J. Winkelmann \corref{email}}

\cortext[email]{\textit{Email of corresponding author:} jwinkelm@tcd.ie}
\address[TCD]{School of Physics, Trinity College Dublin, The University of Dublin. Ireland.}

\author[TCD]{F.F. Dunne}

\author[Lyon]{V.J. Langlois}
\address[Lyon]{Laboratoire de G\'eologie de Lyon, Terre, Plan\'etes, Environnement,
Department of Earth Science, Universit\'e Claude Bernard Lyon 1,
France}

\author[TCD]{M.E. M\"obius}

\author[TCD]{D. Weaire}

\author[TCD]{S. Hutzler}

\date{\textrm{\today}}

\begin{abstract}
Jammed soft matter systems are often modelled as dense packings of overlapping soft spheres, thus ignoring particle deformation.
For 2D (and 3D) soft disks packings, close to the critical packing fraction $\phi_c$, this results in an increase of the average contact number $Z$ with a square root in $\phi-\phi_c$.
Using the program PLAT, we find that in the case of idealised two-dimensional foams, close to the wet limit, $Z$ increases {\em linearly} with $\phi-\phi_c$, where $\phi$ is the gas fraction.
This result is consistent with the different distributions of separations for soft disks and foams at the critical packing fraction.
Thus, 2D foams close to the wet limit are not well described as random packings of soft disks, since bubbles in a foam are deformable and adjust their shape.
This is not captured by overlapping circular disks.

\url{https://doi.org/10.1016/j.colsurfa.2017.03.058}
\end{abstract}


\end{frontmatter}

\section{Introduction}
\begin{figure}[h!]
\centering
\begin{tikzpicture}
\node[anchor=south west,inner sep=0] (image) at (0,0) {
\includegraphics[clip=true, trim=72 172 67 214, width=0.9186\thirdlinewidth]{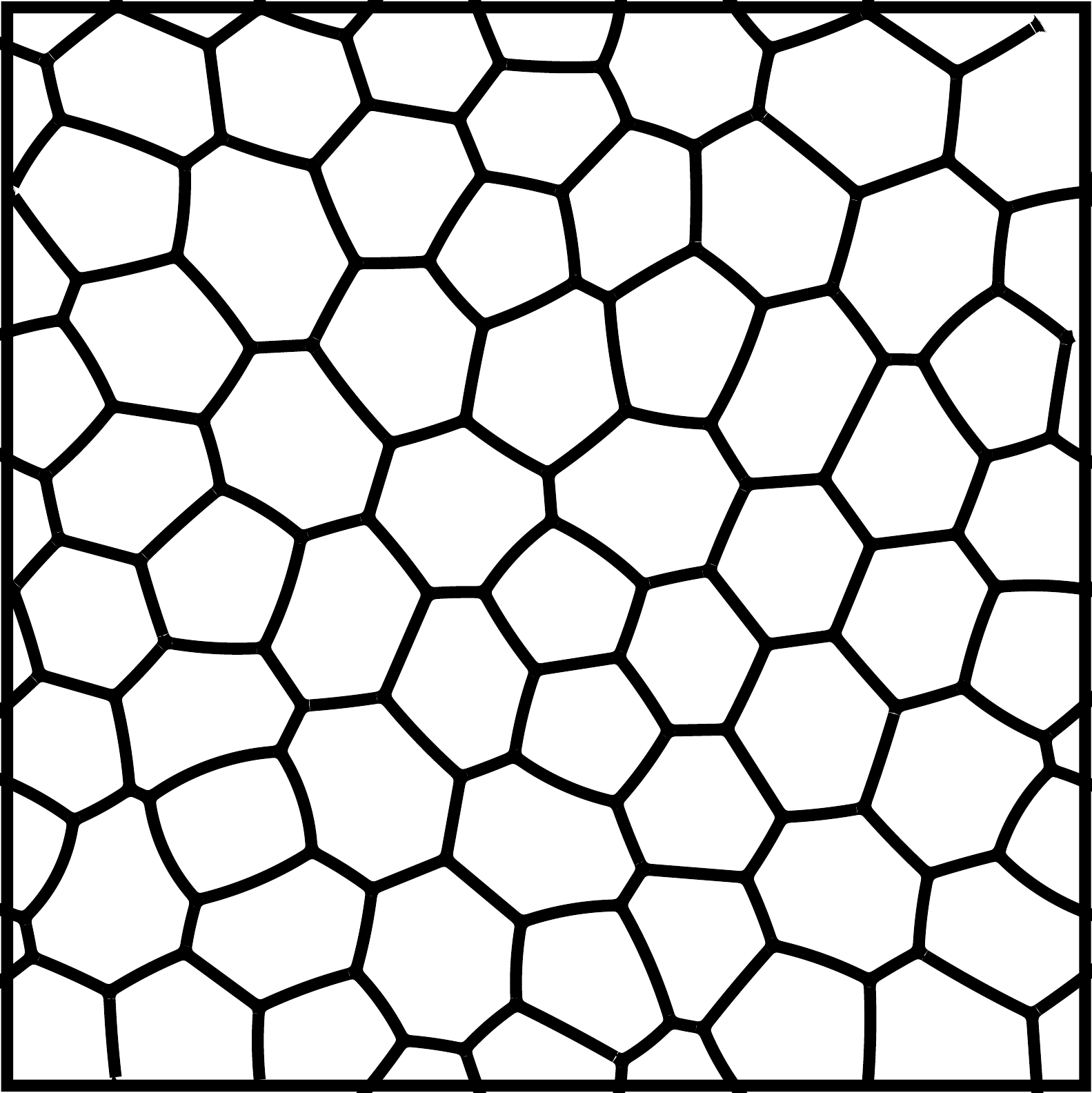}};
\begin{scope}[x={(image.south east)},y={(image.north west)}]
  \node at (0.1, 1.15) {(a)};
\end{scope}
\end{tikzpicture}
\begin{tikzpicture}
\node[anchor=south west,inner sep=0] (image2) at (0,0) {
\includegraphics[clip=true, trim=73 173 83 163, width=0.9688\thirdlinewidth]{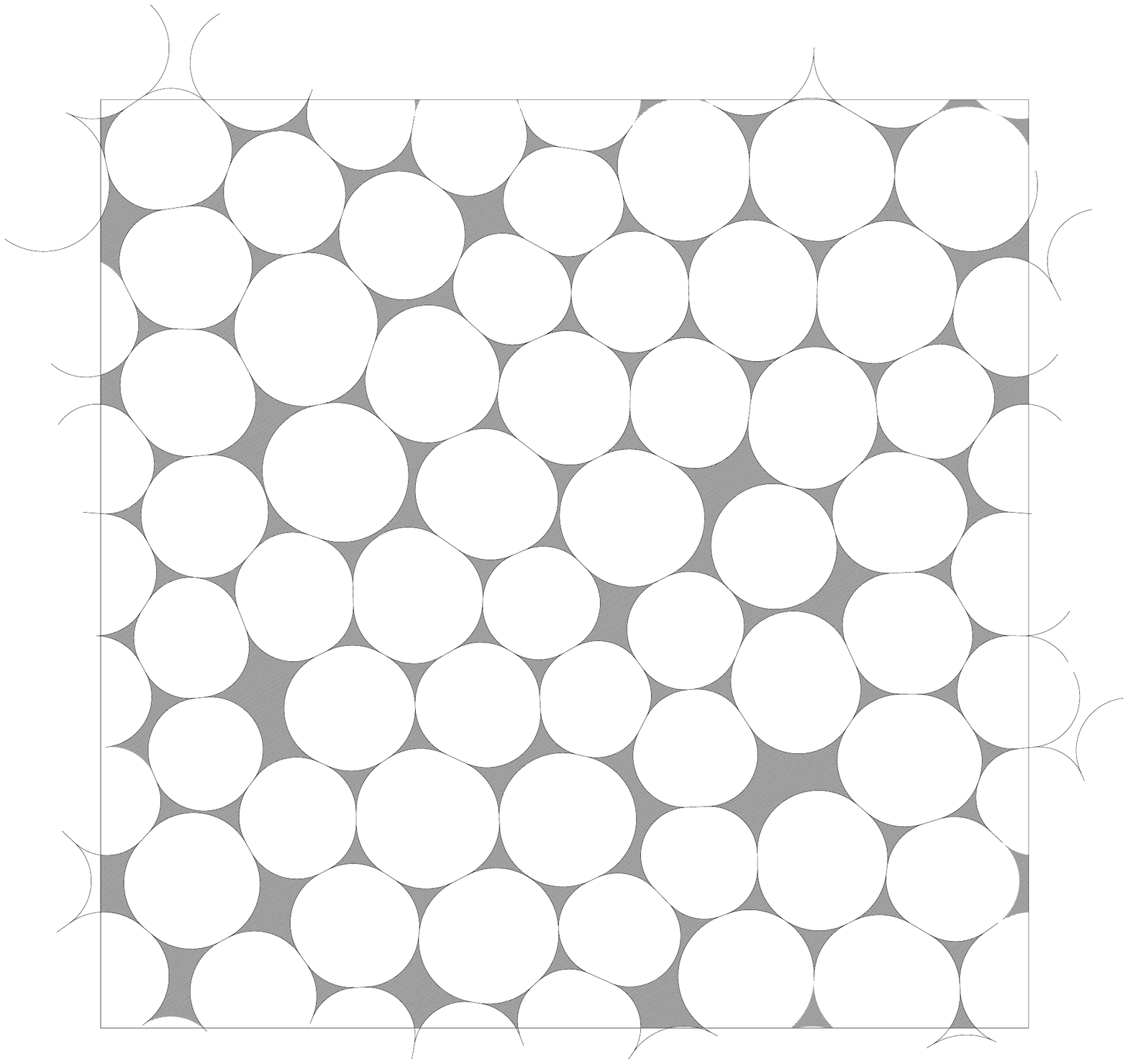}};
\begin{scope}[x={(image.south east)},y={(image.north west)}]
  \node at (0.1, 1.15) {(b)};
\end{scope}
\end{tikzpicture}
\begin{tikzpicture}
\node[anchor=south west,inner sep=0] (image) at (0,0) {
\includegraphics[clip=true, trim=73 173 83 163, width=\thirdlinewidth]{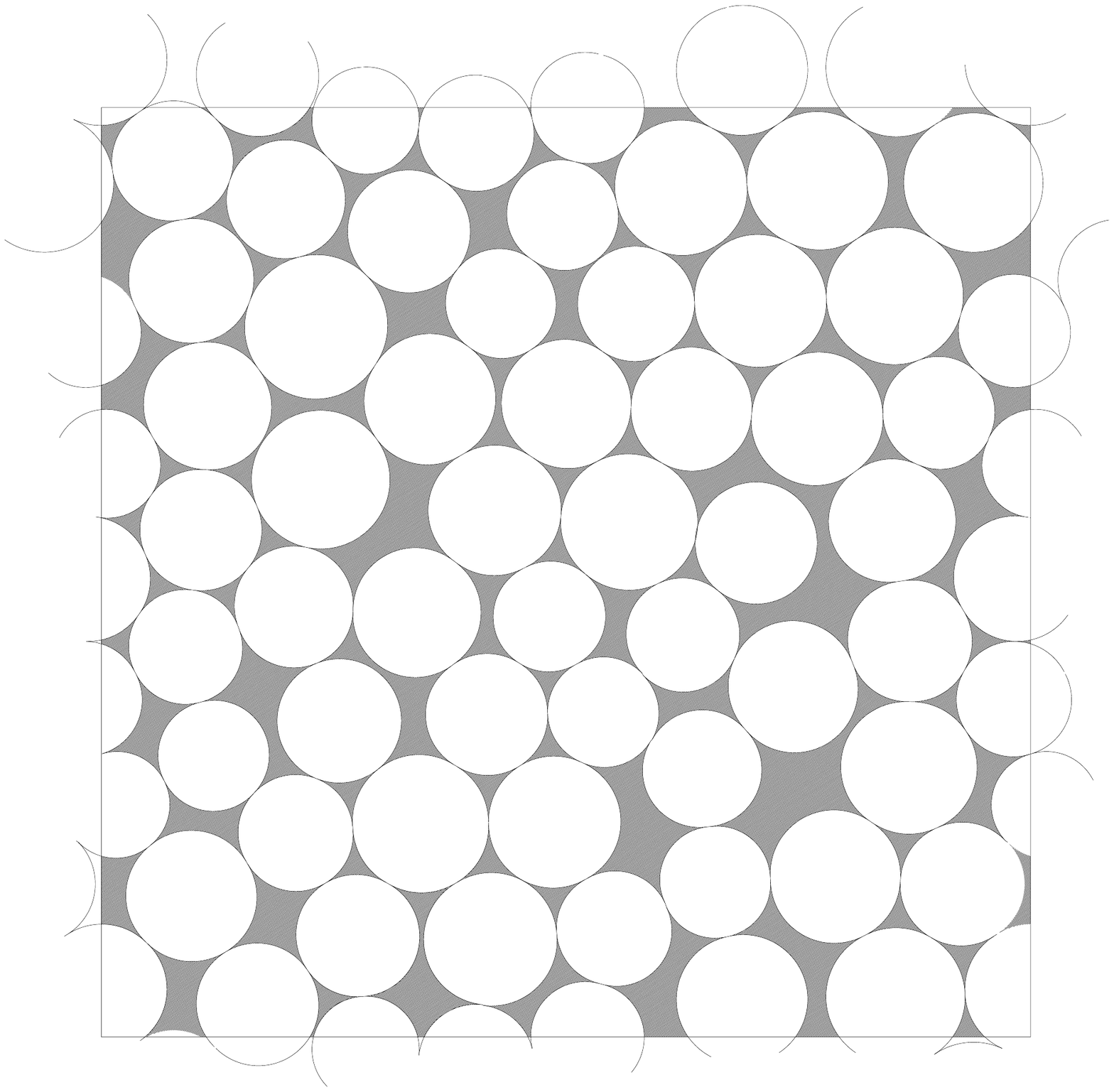}};
\begin{scope}[x={(image.south east)},y={(image.north west)}]
  \node at (0.1, 1.055) {(c)};
\end{scope}
\end{tikzpicture}

\begin{tikzpicture}
    \draw[thick, <->,>=latex](-5,-0.3) -- (3,-0.3);
    \node at (-5,-0.5) {Dry};
    \node at (3,-0.5) {Wet};
\end{tikzpicture}

\begin{tikzpicture}
\node[anchor=south west,inner sep=0] (image) at (0,0) {
\includegraphics[clip=true, trim=210 395 300 310, height=0.9\thirdlinewidth]{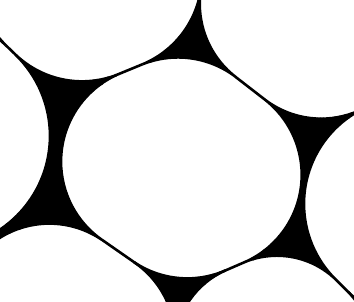}};
\begin{scope}[x={(image.south east)},y={(image.north west)}]
  \node at (0.1, 0.95) {(d)};
  \node[white] (pb) at (0.88, 0.565) {$p_b$};
  \node at (0.5, 0.45) {pressure $p_i$};
  \node at (0.33, 0.87) {$p_j$};
  \node[circle, fill, inner sep =1pt, label={[xshift=-0cm, yshift=0.4cm]0:{$(x_n, y_n)$}}] (vertex) at (0.67, 0.737) {};
  \draw[latex-] (vertex) -- (0.90, 0.81);
  \node[circle, fill, inner sep =1pt] at (0.73, 0.683) {};
\end{scope}
\end{tikzpicture}
\begin{tikzpicture}
\node[anchor=south west,inner sep=0] (image) at (0,0) {
\includegraphics[clip=true, trim=265 195 50 136, height=0.9\thirdlinewidth]{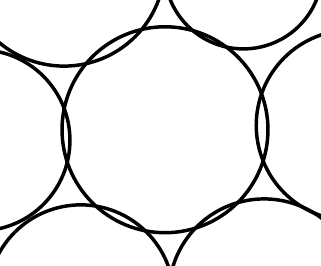}};
\begin{scope}[x={(image.south east)},y={(image.north west)}]
  \node at (0.1, 0.95) {(e)};
\end{scope}
\end{tikzpicture}
\caption{
Sample simulations as obtained with PLAT ((a)--(d)) using periodic boundary conditions of a two-dimensional foam with 60 bubbles at gas fraction $\phi = 0.997$ (a), 0.896 (b) and 0.841 (c).
The bubbles in such a foam are deformed even close to the wet limit, as seen in the example of (d) for $\phi = 0.90$.
In contrast (e) shows an example of overlaps in a soft disk simulation at the same value for $\phi$.
The vertex positions $(x_n, y_n)$ are the coordinates of the point where a Plateau border ends and connects smoothly to a film, separating two bubbles.
}
\label{f:plat_samples}
\end{figure}

In the wet limit a disordered two-dimensional (2D) foam (Fig. \ref{f:plat_samples} (a) -- (c)), as represented by the usual model (incompressible gas and liquid) \cite{foambook1999, cantat2013foams}, assumes the form of a packing of circular disks, as shown in Fig. \ref{f:plat_samples} (c).
Simple arguments, often included in descriptions of jamming of frictionless granular materials, lead to the result that, while local stability requires at least {\em three} neighbours for each disk, overall stability requires {\em four} as an average in 2D \cite{maxwell1864, bennett1972serially, vanHecke2010}.
But how does the average contact number {\em approach} this limiting value, as the wet limit is approached?  

Here we address this question, using the simulation program PLAT \cite{Bolton91,Bolton92} as described below.
It provides a direct and accurate representation of the model (Fig. \ref{f:plat_samples}).

Various experiments for quasi-2D foams \cite{KatgertVanHecke2010} and 2D elastic disks \cite{MajmudarEtal2007}, and simulations with the more approximate {\em soft disk} model \cite{ohern2003jamming} have been in agreement in finding the limiting form for the average contact number $Z$,
\begin{equation}
Z - Z_c \propto (\phi - \phi_c)^{1/2},
\label{e:square-root}
\end{equation}
where $\phi$ is the packing fraction (or gas fraction, in the case of foams) and $\phi_c$ is its critical value;
in the limit of an infinite system the critical contact number is $Z_c = 4$. 

Surprisingly, the result for an ideal 2D foam, simulated with the program by PLAT \cite{Bolton91, Bolton92, PLAT}, which provides a direct and accurate representation of a 2D foam, is different.
It exhibits a {\em linear} increase in the wet limit, $Z - Z_c \propto \phi - \phi_c$.

\begin{figure}[h!]
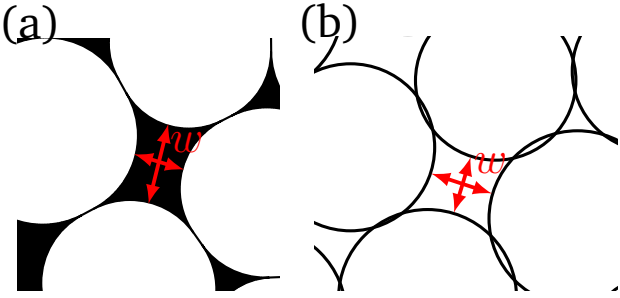

\centering
\begin{tikzpicture}
\node[anchor=south west,inner sep=0] (image) at (0,0) {
\includegraphics[clip=true, trim=110 190 410 513, width=0.39\linewidth]{denseMidThin.pdf}};
\begin{scope}[x={(image.south east)},y={(image.north west)}]
\node at (0.05, 1.05) {\LARGE (a)};
\draw[thick, <->, >=latex, red, line width=0.5mm] (0.5, 0.345) -- node[above right] {\LARGE $w$} (0.575, 0.66);
\draw[thick, <->, >=latex, red, line width=0.5mm] (0.4452, 0.541) -- (0.635, 0.48);
\end{scope}
\end{tikzpicture}
\begin{tikzpicture}
\node[anchor=south west,inner sep=0] (image) at (0,0) {
\includegraphics[clip=true, trim=150 160 160 165, width=0.45\linewidth]{Packing09.pdf}};
\begin{scope}[x={(image.south east)},y={(image.north west)}]
\node at (0.05, 1.05) {\LARGE (b)};
\draw[thick, <->, >=latex, red, line width=0.5mm] (0.46, 0.307) -- node[above right] {\LARGE $w$} (0.515, 0.525);
\draw[thick, <->, >=latex, red, line width=0.5mm] (0.385, 0.46) -- (0.585, 0.382);
\end{scope}
\end{tikzpicture}
\caption{
An Illustration of the separation $w$ between two bubbles (a) and soft disks (b).
The separation is defined as the shortest distance between two bubble arcs/disk edges.
Its distribution $f(w)$ is connected to $Z - Z_c$ via an integration \cite{vanHecke2010, ohern2003jamming} (see also eqn. \eqref{eq:Zintegral}).
}

\label{fig:separations}
\end{figure}

This result is consistent with the distribution of separations \cite{SiemensVanHecke2010} $f(w)$ for the 2D foam, which is connected to $Z - Z_c$ via an integration \cite{vanHecke2010, ohern2003jamming}.
This separation $w$ is defined as the shortest distance between two bubbles/disk edges (see Fig. \ref{fig:separations}).
While $f(w)$ for the soft disks exhibits a square root divergence, it reaches a finite limiting value for the foam in the limit of $w \rightarrow 0$.

\section{Computer simulation of 2D foams}

The results for the average contact number $Z(\phi)$ presented below were produced by the PLAT simulation code from \cite{PLAT} as described in \cite{Bolton91, Bolton92, BoltonWeaire90}.

It is a software for the simulation of random 2D foam \cite{Bolton91, Bolton92, BoltonWeaire90, PLAT} which is not based on an energy minimisation routine, but instead directly implements Plateau's laws for a 2D foam by modelling the films and liquid-gas interfaces as circular arcs, constrained to meet smoothly at vertices, see Fig. \ref{f:plat_samples} (d).
The radius of curvature $r$ of each arc is determined by the Laplace law.

For a film this law is $p_i - p_j = 2 \gamma / r$, where $p_i$ und $p_j$ are the pressures in the two adjacent bubbles and $\gamma$ is the surface tension.
For a liquid-gas interface $p_i - p_b = \gamma / r$, where $p_b$ is the pressure in the Plateau border, set equal in all Plateau borders.

The samples were generated as (nearly) dry foams by standard procedures \cite{Bolton91, Bolton92, DunneEtal2016}:
A random Delauney tessellation is used to compute a Voronoi network.
This is then converted to a (as yet unequilibrated) dry foam by decorating its vertices with small three-sided Plateau borders.
The equilibration process of the decorated Voronoi network consists of adjusting cell pressure and the vertex positions $(x_n, y_n)$ under the constraints of smoothly meeting arcs and area conservation for each bubble.
Equilibrium is reached when the change in vertex positions is small. 

A progressive decrease in steps of $\Delta \phi = 0.001$ in gas fraction was imposed and the system was equilibrated at each step.
Decreases in gas fraction are performed by proportionally reducing bubble areas.
The bubble radius distribution of the sample, which is calculated from bubble cell areas, follows a lognormal distribution with a standard deviation $\Delta R / \langle R \rangle \approx 0.07$.
More details of the protocol for sample preparation are given in \cite{DunneEtal2016}.

Note that PLAT is currently the only simulation that can simulate a wet foam with zero contact angle between two liquid interfaces.
The Surface Evolver \cite{brakke1992}, the standard software to simulate 2D and 3D foams, requires finite contact angles with consequences that are currently being examined \cite{kraynik}.

As in its earlier application \cite{Hutzler95}, PLAT was found to be susceptible to a lack of convergence close to $\phi_c$, which has not yet been eliminated.
In the present case, this was mitigated by using a fairly small system (with periodic boundary conditions), consisting of $60$ bubbles, as in Fig. \ref{f:plat_samples} (a) to (c).
Results from $\num{600000}$ independent simulations were combined to compute the variation of $Z(\phi)$.
Finite size effects were taken into account when estimating the critical packing fraction $\phi_c$, as detailed below.
We believe this procedure to be reliable for present purposes, although there is a slight possibility of undesirable bias in the surviving runs close to the wet limit.

As a standard procedure \cite{KatgertVanHecke2010, ohern2003jamming}, rattlers, which are bubbles with less than three contacts, were excluded in our analysis.
These do not contribute to the connected network and are mechanically unstable bubbles, which can be removed without changing the packing.
(In the wet limit, less than $\SI{4}{\percent}$ of all bubbles were rattlers.)

For a comparison with the soft disk model, random packings with similar conditions (same polydispersity, same sample preparation protocoll) as in PLAT were created using conjugate gradient energy minimisation \cite{numpy}.
The average for $Z(\phi)$ excluding rattlers were taken over $\num{20000}$ independent simulations.

In analysing our results we need to take into account a small finite-size correction.
In an infinite disordered packing of disks the critical packing fraction $\phi_c$ is associated with a contact number $Z=4$, according to arguments based on counting constraints \cite{maxwell1864,bennett1972serially, vanHeckePRL}.
 In the case of our finite system with periodic boundaries the critical value of the contact number is given by $Z_c = 4(1-\nicefrac{1}{N})$, where $N$ is the number of bubbles; $N=60$ in our case thus results in $Z_c = 3.933$.
This relation is obtained from matching the number of degrees of freedom, $2N$ for a two-dimensional packing, with the number of constraints, due to the $Z N/2$ contacts.
However, in a periodic system we can fix one bubble without loss of generality, leaving only $N-1$ bubbles free to under\-go translational motion.

\section{The variation of $Z(\phi)$ for 2D foams}

\begin{figure}[h!!]
\centering
\begin{tikzpicture}
\node[anchor=south west,inner sep=0] (image) at (0,0) {
\includegraphics[width=\linewidth]{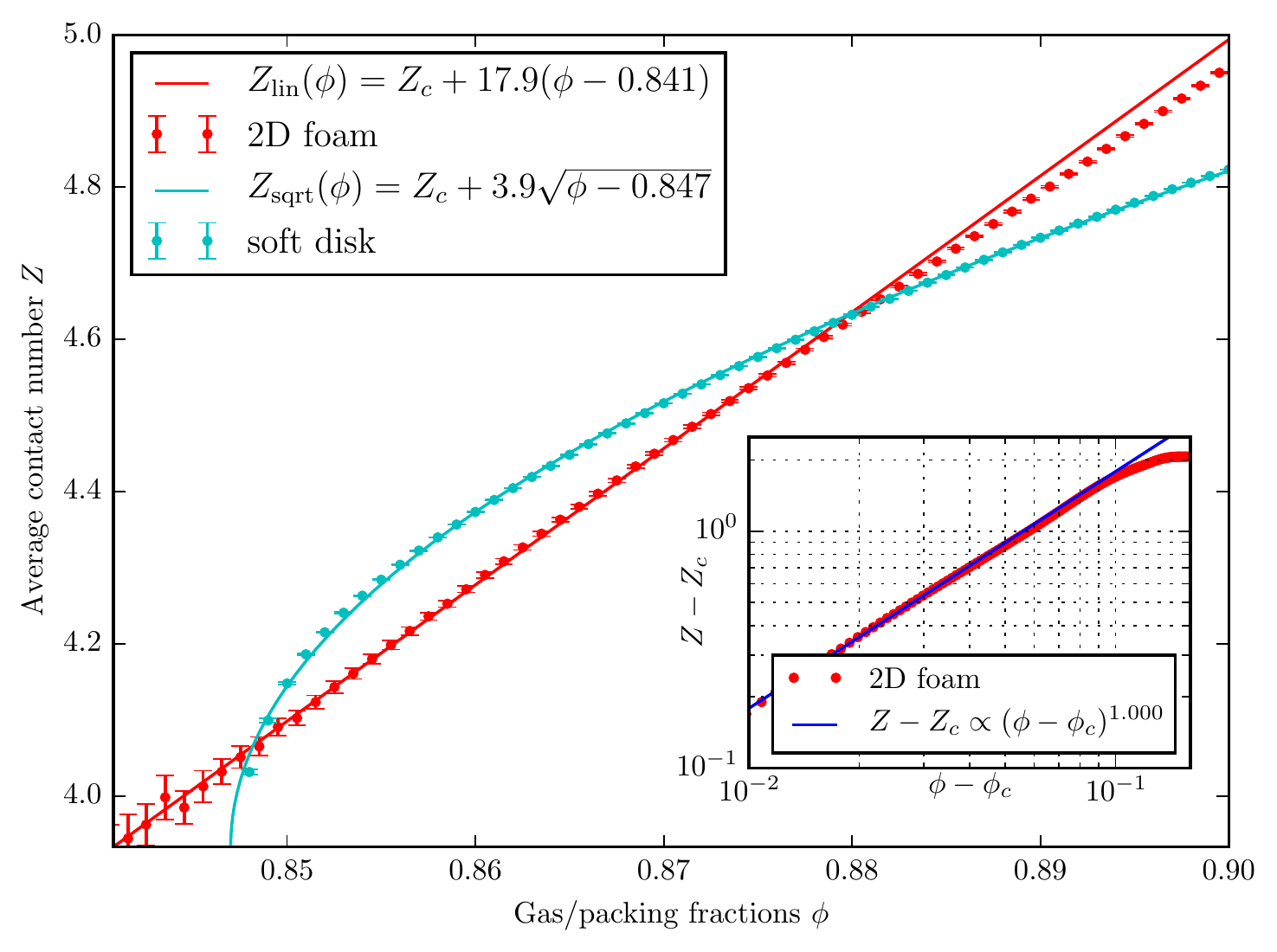}};
\begin{scope}[x={(image.south east)},y={(image.north west)}]
  \node at (0.08, 0.12) {\Large $Z_c$};
\end{scope}
\end{tikzpicture}
\caption{
For 2D foams close to the critical gas fraction the average number of contacts $Z$ without rattlers was found to vary linearly with $\phi-\phi_c$ (red data points).
The average was taken over $\num{600000}$ independent simulations with $60$ bubbles.
A linear fit (solid red line) in the displayed range gave a slope of $k_{\mathrm{f}} = 17.9 \pm 0.1$ and a critical gas fraction of $\phi_c = 0.841 \pm 0.001$.
In the wet limit (at $\phi_c$), $Z_c$ is given by $Z_c = 4(1-\nicefrac{1}{N})$ due to finite size effects.
This results in $Z_c = 3.933$ for $N=60$ bubbles.
For comparison, $Z(\phi)$ is also plotted for a soft disk systems ($N = 60$ with $\num{20000}$ realisations), which shows the mentioned square-root scaling.
Inset: Double-logarithmic scale for $Z - Z_c$ vs. gas/packing fractions $\phi - \phi_c$ up to $\phi = 1$.
By fitting a linear function (solid line), the $\phi_c$ which gives the best linear relationship is obtained as $\phi_c = 0.841 \pm 0.001$.
}
\label{f:z_phi_foam}
\end{figure}

In order to investigate the variation of $Z(\phi)$ close to $\phi_c$, and the value of $\phi_c$ itself, we plotted $\log(Z(\phi)-Z_c)$ vs. $\log(\phi - \phi_c)$, varying $\phi_c$ to obtain the value which gives the best {\em linear} relationship between these quantities (see also inset plot of Fig. \ref{f:z_phi_foam}).
In this way, the critical gas fraction was found to be $\phi_c~=~0.841 \pm 0.001$, and the slope was $1.000 \pm 0.004$ in the logarithmic plot.

The conclusion is therefore that $Z$ approaches $Z_c$ \emph{linearly}, i.e. $(Z-Z_c) \sim (\phi - \phi_c)$ as plotted in Fig. \ref{f:z_phi_foam}.
Appropriately, fitting
\begin{equation}
Z  = Z_c +  k_{\mathrm{f}} (\phi - \phi_c),
\label{e:linear}
\end{equation}
with $Z_c = 4-\nicefrac{1}{15}$ gives $k_{\mathrm{f}} = 17.9 \pm 0.1$ and a critical gas fraction of $\phi_c = 0.841 \pm 0.001$.
In a different approach, by looking at the excess energy, we obtained $\phi_c = 0.839~\pm~0.001$  \cite{DunneEtal2016} for the same system.

The value of $\phi_c$ is consistent with previous experimental and numerical results, obtained for example from measurements of packings of bidisperse hard disks\cite{bideau1984compacity}, bidisperse elastic disks\cite{MajmudarEtal2007}, polydisperse hard disks\cite{bideau1984compacity}, experimental data for (quasi) two-dimensional foams\cite{KatgertVanHecke2010}, and computer simulations of polydisperse soft disk packings \cite{Durian95}.
In the dry limit at $\phi = 1$, the PLAT simulation leads to $Z = 6$, which is the expected average contact number \cite{foambook1999}.
This is not the case for the soft disk model.

Our findings are also consistent with cruder estimates from previous PLAT simulations \cite{BoltonWeaire90, Diplomarbeit} and simulations using a hybrid lattice gas model \cite{SunHutzler2004}.


\section{Discussion of previous results for $Z(\phi)$}
The linear increase of the average contact number with gas fraction, close to the wet limit, eqn. \eqref{e:linear}, is unexpected, since it is at odds with many previous findings from computation, theory, and experiment.
As an illustration we plot in figure \ref{f:z_phi_foam} also results from soft disk systems with the same radius polydispersity as our 2D foam.

Thus, before presenting further results supporting our results, we want to discuss the contradiction with previous results and how to resolve it.

At first there might seem to exist an incontrovertible weight of evidence for the square-root scaling, eqn.~\eqref{e:square-root}, but this is not the case for the 2D foam.
 We discuss the two strands of contrary evidence in turn.
These are, firstly, results from the soft-disk model, and secondly, experimental data for bidisperse 2D foams.

The discovery of the square root scaling for $Z(\phi)$ appears to date back to the work of Durian using the so-called Bubble Model \cite{Durian95}.
Durian developed this model primarily to investigate the rheological properties of foams, of which it indeed provides a good overall description \cite{LangloisEtal08}.
Two-dimensional bubbles are approximated as disks, subject to repulsive forces when they overlap.

The same square-root scaling for $Z(\phi)$ was also found in computer simulations of packings of three-dimensional soft spheres \cite{ohern2002}, a system which has since been called the ```Ising model' for jamming'' \cite{vanHecke2010}.

If one describes foams in the wet limit as packings of disks (or spheres), then it is tempting to extend this analogy also to the functional relationship for $Z(\phi)$ and thus expect the same square-root relationship in lowest order.
However, Surface Evolver simulations have shown, while the energy is harmonic in 2D, the bubble-bubble interactions are not pairwise-add\-itive \cite{Lacasse96}.
That is, the model of interaction that lies at the heart of the soft disk model does not represent realistic bubble-bubble interactions.
One should therefore treat this prediction with some caution.

Experimental evidence of the square-root scaling, as found from measurements of two-dimensional photoelastic disks under compression \cite{MajmudarEtal2007}, is in agreement with the prediction of the bubble-model, which one might expect to be applicable in this case, at least for qualitative purposes.

Let us now turn to the second strand of contrary evidence by examining further experimental results which bear directly on 2D foams.

Katgert and van Hecke \cite{KatgertVanHecke2010} performed experiments with disordered rafts of \textit{bidisperse} bubbles beneath a glass plate.
The distance between plate and liquid surface was varied to obtain foams at different values of gas fractions.
The concept of a gas fraction is not well defined for such {\em quasi}-2D bubble rafts, in particular in the wet limit where the gap between covering plate and liquid interface is similar to the bubble extension parallel to the plate.
For this reason Katgert and van Hecke \cite{KatgertVanHecke2010} proceeded by imaging their rafts from the top to obtain an {\em area} gas fraction.
Based on their analysis Katgert and van Hecke established $Z - Z_c \propto (\phi - \phi_c)^{\alpha}$, with exponent $\alpha \simeq 0.70$, $Z_c$ close to 4, and $\phi_c$ close to $0.84$ \cite{KatgertVanHecke2010}.
Due to the problem in defining a gas fraction for such a quasi-2D experiment, and in identifying contacting bubbles, we do not think that these experimental results can be taken to contradict our PLAT findings, even though Katgert and van Hecke describe their wet foams as consisting of ``soft frictionless disks''.

For 3D foams our results suggest also a deviation from the square root scaling in $Z(\phi)$, since we conjecture the reason for the deviation in the 2D case to be the model of interaction.
However, the scaling does not have to be linear.
Apart from the non-pairwise interaction, the energy for the 3D bubble-bubble interaction is also not harmonic.
It scales with the form $f^2 \ln(1 / f)$, first predicted by Morse and Witten, where $f$ is the force exerted between droplets \cite{MorseWitten, Lacasse96, ZCone2014}.

However, similar to the 2D case, evidence for the square root scaling seems to be indisputable at first glance.
Experiments from Jorjadze \textit{et al.} \cite{Jorjadze2013} with droplet emulsion in 3D show a good agreement with the square root increase in $Z(\phi)$.
But, as in the experiments of Katgert and van Hecke the identification of contacting bubbles and the definition of a gas fraction is not straight forward.
Jorjadze \textit{et al.} reconstructed the droplets as overlapping spheres and defined contacts as overlaps.
The gas fraction is then the spherical volume reduced by the overlaps.
Thus, it cannot be ruled out that this procedure contains a bias towards the square root scaling of $Z(\phi)$ as in the soft disk model.

The distribution of contacts in a packing can be predicted via the granocentric model \cite{Clusel2009} which has recently been extended to 2D cellular structures \cite{Miklius2012} and 2D packings of discs \cite{Cathal2013}.
However, this model cannot predict the variation of $Z$ with $\phi$ in packings as it only applies to the wet limit (or jamming point).

\section{Link between $Z(\phi)$ and the radial density function $g(r)$}
For soft disk packings it has been argued that the square root scaling of $Z$ as seen in \eqref{e:square-root} is connected with the variation of the radial density function $g(r)$ via an integration \cite{vanHecke2010, ohern2003jamming, Wyart}, although the validity of this argument is still under discussion \cite{Charbonneau}.

The radial \textit{distribution} function $R(r)$ is defined as the probability to find a particle a given distance $r$ away from another particle.
In 2D the radial \textit{density} function is given by $g(r) = \frac{1}{2\pi r} R(r)$.
From simulations of 3D monodisperse soft spheres with diameter $D = 1$ close to the jamming transition the behaviour of $g(r)$ is found to be divergent, according to the power law
\begin{equation}
g(r) = \frac{c_{\mathrm{d}}}{\sqrt{r - 1}}\,,
\label{eq:g(r)}
\end{equation}
where $c_{\mathrm{d}}$ is a constant \cite{ohern2003jamming}.
A similar divergence can be found in 2D polydisperse systems, when the radial density function is rescaled to $g(\xi)$ with the rescaled interparticle distance $\xi~=~r~/~(R_i~+~R_j)$, where $R_i$ and $R_j$ are the radii of two disks with distance $r$ apart \cite{vanHecke2010}.
Using an affine Ansatz (see below), integrating $g(r)$ over $r$ then results in the square root scaling for $Z(\phi)$ of eqn. (\ref{e:square-root}) \cite{vanHecke2010, ohern2003jamming, Wyart}.
\label{section:rdf}

\section{Distribution of separation $f(w)$ for 2D foams and soft disk systems}
\label{s:contacts}
For 2D foams such an argument involving $g(\xi)$ is not straightforward to develop, since bubbles are deformable and only have well-defined centres in the wet limit (at $\phi_c$) where they are circular.
For this reason we will in the following consider a different approach, which involves a distribution of separations $f(w)$ between bubbles (or disks), as in the work of Siemens and van Hecke \cite{SiemensVanHecke2010}.
Here, the separation $w$ is the shortest distance between two bubble arcs/disk edges (see Fig \ref{fig:separations}).
For the soft disk system, this separation is then related to their distance by their radii, $r = w + R_i + R_j$.   
For the soft disk system $f(w)$ is identical to $g(\xi)$ close to the divergence, when shifted by the average disk diameter $D$, thus $g(\xi - D) = f(w)$.

\begin{figure}[h!]
\begin{tikzpicture}
\node[anchor=south west,inner sep=0] (image) at (0,0) {
\includegraphics[width=\linewidth]{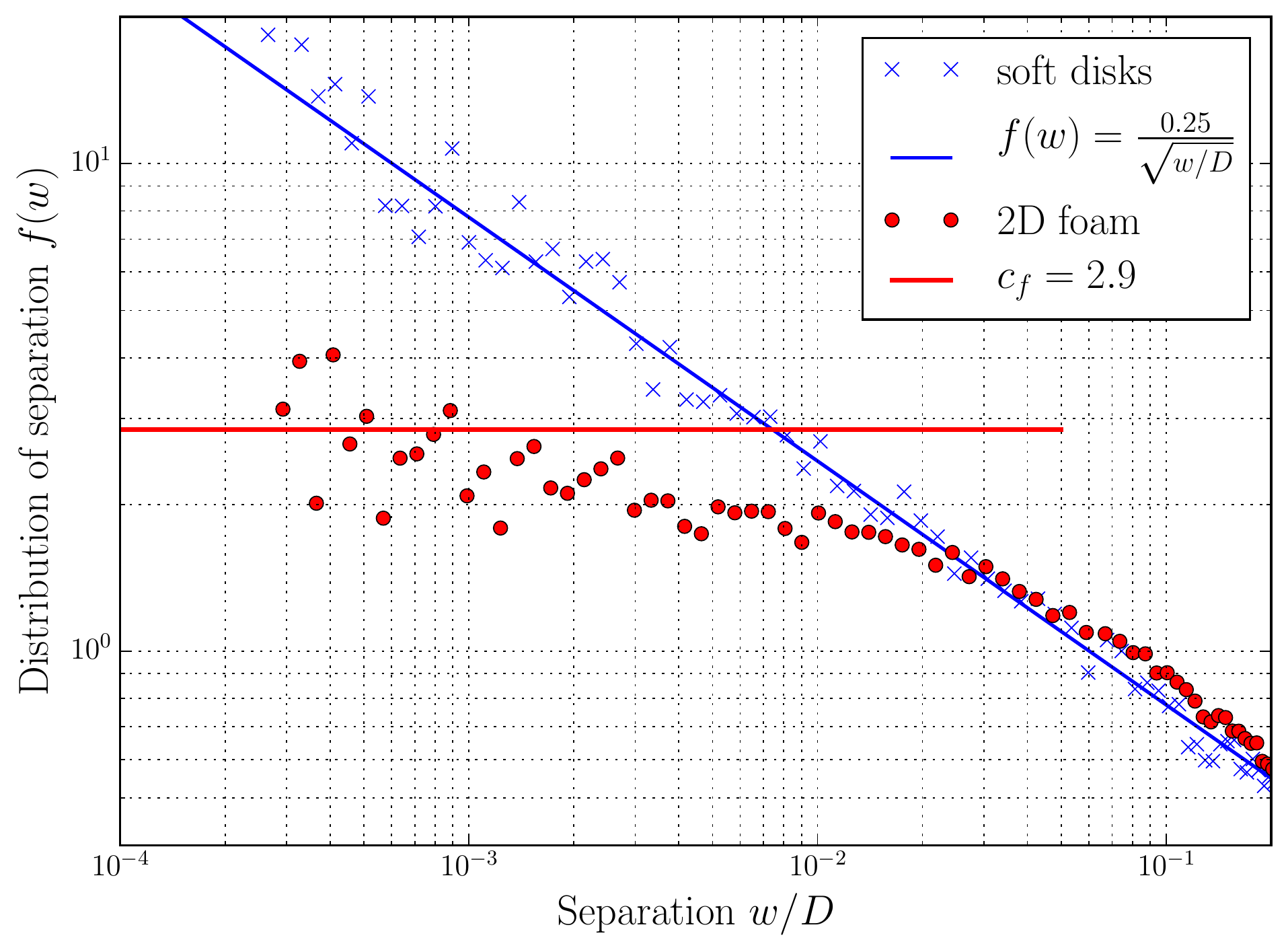}};
\begin{scope}[x={(image.south east)},y={(image.north west)}]
  \node[text width=3cm] (foams) at (0.5, 0.27) {finite limiting value for foams};
  \node[text width=2cm] (softdisks) at (0.55, 0.85) {divergence for soft disks};
  \draw[-latex] (softdisks) -- (0.275, 0.85);
  \draw[-latex] (foams) -- (0.5, 0.585);
\end{scope}
\end{tikzpicture}
\caption{
Distribution of separation $f(w)$ for 2D foam (red circles) and 2D disk packing (blue crosses) at a similar average contact number $Z_{\mathrm{SD}} = 4.07 \pm 0.01$ for soft disks and $Z_{\text{foam}} = 4.06 \pm 0.01$ for the 2D foam ($D$: average bubble/disk diameter).
The data shown presents averages obtained from 1379 packings, each containing 60 bubbles or disks.
In the case of foams, the finite value at $f(w)$ in the limit of $w/D \to 0$ is consistent with the observed linear increase of the average contact number $Z$, according to the approximate argument, given in the text.
The decay of $f(w) \propto (w/D)^{-1/2}$ in the same limit in the case of the disk packings is consistent with the square root increase of the average contact number $Z$.
}
\label{f:distributionpbwdith2}
\end{figure}
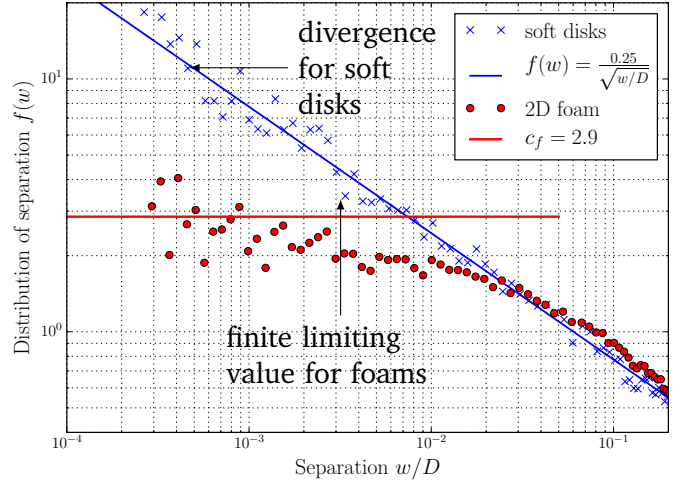

Fig. \ref{f:distributionpbwdith2} shows the distribution $f(w)$ for both foams and packings of soft disks with the same system size ($N = 60$) and area polydispersity.
The difference between our results for simulated 2D foams and 2D disk packings is striking.
Whereas in the case of disks, $f(w)$ diverges in the limit 
\begin{equation}
w / D \to 0 \quad \text{as} \quad f(w) = \frac{c_{\mathrm{d}}}{\sqrt{w / D}}
\label{e:f(w)}
\end{equation}
 as expected from the divergence of $g(\xi)$ with $c_{\mathrm{d}} = 0.25 \pm 0.01$, for the 2D foams a finite limiting value $c_{\mathrm{f}} = 2.9 \pm 0.7$ is reached in this limit.
Only at values of $w / D \gtrsim 10^{-2}$, $f(w)$ is the same for both foams and soft disks; see Fig. \ref{f:distributionpbwdith2}.

Let us now consider the compression of a two-dim\-ens\-ion\-al, polydisperse foam/disk sample of initial gas/ packing fraction $\phi_c$ to a final value of $\phi~>~\phi_c$.
The fractional compression $\Delta \epsilon$ is given by $\Delta \epsilon =  (\phi-\phi_c)/(2\phi_c)$, where $\Delta \epsilon$ is considered to be small.

We can estimate $Z(\phi)$ for the case of an {\em affine compression} from $f(w)$.
In this case the deformation of the sample will lead to an increase in contact number due to bubbles coming together that initially, i.e. in the wet limit (at $\phi_c$), were closest to each other.
For an affine deformation the fractional compression can be expressed as $\Delta \epsilon \approx \Delta w / D$.
Thus, the average number of contacts in 2D can be estimated by integrating $\rho f(w)$ over a radial shell up to $D \Delta \epsilon$, where $\rho = \frac{4 \phi_c}{\pi D^2}$ is the particle number density,
\begin{equation}
Z(\phi) - Z_c = 2 \pi \rho \int_0^{D \Delta \epsilon} \dx w f(w) (D + w)
\label{eq:Zintegral}
\end{equation}

When inserting the power law expression from eqn. \eqref{e:f(w)} into \eqref{eq:Zintegral}, we obtain for the soft disk simulation
\begin{align}
Z(\phi) - Z_c =& \sqrt{128 \phi_c} c_{\mathrm{d}} \sqrt{\phi - \phi_c} + \mathcal{O}\left(\sqrt{\phi - \phi_c}^3\right) \nonumber \\
{}\approx& (2.6 \pm 0.1) \sqrt{\phi - \phi_c}\,,
\end{align}
where we neglected terms of higher order in $\phi - \phi_c$.
For $\phi_c$ the value $0.841 \pm 0.002$ was used \cite{Durian95}.

Fitting the soft disk data for $N = 60$ to a square root function, $Z - Z_c = k_d \sqrt{\phi - \phi_c}$ for all $Z < 4$ gives $k_d = 3.86 \pm 0.01$ and $\phi_c = 0.847 \pm 0.001$.

For the 2D foam simulation, the finite limiting value $c_{\mathrm{f}}$ can be inserted for $f(w)$ in the limit $w / D \rightarrow 0$ in eqn. \eqref{eq:Zintegral}.
By integrating we then obtain for $Z(\phi)$
\begin{align}
Z(\phi) - Z_c =& 4 c_{\mathrm{f}} (\phi - \phi_c) + \mathcal{O} \left( (\phi - \phi_c)^2 \right) \nonumber\\
{}\approx& (11.6 \pm 2.8) (\phi - \phi_c)\,.
\end{align}
Again, we neglected terms of higher order in $\phi - \phi_c$.

Qualitatively both estimations are in accord with expectations, although the apparent numerical discrepancy in the prefactor remains to be resolved.
In both cases the prefactors are underestimated when obtained from our data for soft disk/bubble separations.

\begin{table}[h!]
\addtolength{\tabcolsep}{-0.15cm}  
\centering
\footnotesize{
\begin{tabular}{l l l l}
\toprule
\multirow{2}{*}{} & \multirow{2}{1.4cm}{$\displaystyle{\lim_{w \to 0} f(w)}$} & \multicolumn{2}{c}{$Z(\phi) - Z_c$} \\\cline{3-4}
& & Computed via $f(w)$ & Direct calculation\\\midrule
2D foam: & $2.9 \pm 0.7$ & $(12 \pm 3) (\phi - \phi_c)$ & $(18.1 \pm 0.1) (\phi - \phi_c)$ \\\midrule
soft disks: & $\frac{0.25\pm0.01}{\sqrt{w / D}}$ & $(2.6 \pm 0.1) \sqrt{\phi - \phi_c}$ & $(3.86 \pm 0.01) \sqrt{\phi - \phi_c}$ \\\bottomrule
\end{tabular}
}
\caption{
A summary of our results for 2D foams and soft disks.
The functional form for $Z(\phi)$ can be obtained from the distribution of separation $f(w)$ for both 2D foams and soft disks, the numerical prefactor is underestimated for both by a factor of $3 / 2$.
}
\label{tab:summary}
\addtolength{\tabcolsep}{0.15cm}  
\end{table}

Table \ref{tab:summary} summarises all results that we found to differ in 2D foams and soft disks.
It demonstrates that the linear variation of $Z$ close to $\phi_c$ is consistent with the distribution of separation found in wet foams.
However, this is still short of a full explanation of the asymptotic properties of the wet limit.

\section{Conclusions}
The variation of $Z$ as a function of gas fraction was one of the first problems that were tentatively addressed with the PLAT software, as soon as it was developed in the early 1990s.
The very limited data sets available at the time ($\phi \ge 0.875$, 100 cells \cite{BoltonWeaire90}, 530 cells \cite{Diplomarbeit}) showed that a linear extrapolation of the data leads to $Z=4$ at $\phi_c \simeq 0.84$ \cite{bideau1984compacity}. 
However, later simulations using a lattice gas model for foams also showed a linear variation of $Z$ very close to $\phi_c$, but this data was based on an even smaller sample of only 30 bubbles \cite{SunHutzler2004}.

The success of Durian's bubble model \cite{Durian95, Durian97}  in reproducing the Herschel--Bulkley type rheology that is associated with emulsions and foams \cite{LangloisEtal08}, and its ease in simulating packings of 10000 or more bubbles, led to it being treated as the most practical model for simulations of 2D foams in general.
Its square-root variation of $Z$ with gas fraction away from $\phi_c$ was thus expected to also hold for 2D foams.
Here we have shown, based on a large amount of new data, that this is not the case.
For 2D foams we find that the average contact number varies linearly in this limit.

The reason for this differing behaviour must ultimately lie in the different contributions that disk or bubble contacts make to the total energy of the packing.
In a foam the energy per bubble {\em per contact} increases with the number of contacts \cite{Lacasse96}.
Energy minimisation might thus lead to the reduction in the number of contacts in the wet limit compared to disk packings.

In summary, we showed that the disordered structure of a polydisperse 2D foam is significantly different compared to a soft disk packing with the same polydispersity as evidenced by the different $Z(\phi)$ and corresponding distribution of separations.
This is due to the deformation of the bubbles, which is absent in the soft disk model, and the lack of pairwise interactions.
While this study only focussed on a 2D foam system, similar deviations are likely for other 2D jammed systems with soft, deformable particles.
The relevance to 3D packings of soft particles, such as emulsions, biological cells \cite{biocells, BioHilgenfeldt} and microgel particles \cite{microgel} remains to be examined.

\section*{Acknowledgments}
We would like to thank F. Bolton for updating the PLAT software and D.~McDermott for carrying out some of the initial numerical analysis of the distribution of near contacts.
Research supported in part by a research grant from Science Foundation Ireland (SFI) under grant number 13/IA/1926 and from an Irish Research Council Postgraduate Scholarship (project ID GOIPG/2015/1998).
 We also acknowledge the support of the MPNS COST Actions MP1106 `Smart and green interfaces' and MP1305 `Flowing matter' and the European Space Agency ESA MAP Metalfoam (AO-99-075) and Soft Matter Dynamics (contract:\newline4000115113).


\end{document}